\documentstyle[aps,eqsecnum,preprint]{revtex}
\tighten 

\begin{document}

\draft 

\title{Extension of Kohn-Sham theory to excited states by means of an
off-diagonal density array}

\author{Abraham Klein
\thanks{abklein1@home.com}} 
\address{Department of Physics, University of Pennsylvania, Philadelphia,
PA 19104-6396}
\author{Reiner M.\ Dreizler
\thanks{dreizler@th.physik.uni-frankfurt.de}}
\address{Institut f\"ur Theoretische Physik, Universit\"at Frankfurt,
D-60054 Frankfurt}

\date{\today}

\maketitle

\begin{abstract}

Early work extending the Kohn-Sham theory to excited states 
was based on replacing the study of the ground-state energy as a functional
of the ground-state density by a study of an ensemble average of the
Hamiltonian as a functional of the corresponding average density.
We suggest and develop an alternative to this description of excited states
that utilizes the matrix of the density operator taken between any two states
of the included space.  Such an approach provides
 more detailed information
about the states included, for example,
transition probabilities between discrete states of local one-body operators.
The new theory is also based
on a variational principle for the trace of the Hamiltonian over the space of
states that we wish to describe viewed, however, as a functional
of the associated array of matrix elements of the density. It finds expression
in a matrix version of
Kohn-Sham theory.  To illustrate the formalism, we study a
suitably defined weak-coupling limit, which is our equivalent
of the linear response approximation.  On this basis, we derive an eigenvalue equation that has the same form as an
equation derived directly from the time-dependent Kohn-Sham equation and 
applied recently with considerable success to molecular excitations.  We provide an independent proof,
within the defined approximations, that the eigenvalues can be interpreted as true excitation energies,

\end{abstract}

\pacs{31.15.Ew, 32.15.Ne, 31.15.Pf}

\section{Introduction}
 
Density functional theory (DFT) was designed originally as a theory of the
ground-state of a many-particle system
\cite{KD,KS,PY,DG,HM}.  For an extension to include the calculation
of excitation energies, several lines of thought have been developed.  The
earliest was based on a minimum principle \cite{tr1,tr2} for the trace of
the Hamiltonian over a set of the lowest-energy eigenstates of the system.
This theory has been extended to a suitably weighted sum over the same set
of eigenstates \cite{tr3}.  The expanded version of the Hohenberg-Kohn
theorem, in either case,  is that the average energy
is a unique functional of the corresponding
average density.  Excitation energies are (essentially) obtained by taking
differences between averages over almost overlapping sets.  This approach has not
been developed beyond the cited work.

Recently, considerable attention has been focused on the development of
other methods for studying excitation energies.
One approach is based on
time-dependent density functional theory (TDDFT)
\cite{tdf1,tdf2,tdf3,tdf4,tdf5,tdf6}.  Here one studies
the linear response of the time-dependent density to a time-dependent external field.  The Fourier transform of the
susceptibility (density-density correlation function), which
is the essential ingredient for the calculation of dynamic polarizabilities,
has poles at the true eigenstates of the system. 

From these elements two different formalisms for the calculation
of energies and polarizabilities have been derived, termed 
density based and density-matrix based coupled Kohn-Sham schemes
(CKS) in a recent publication \cite{GHR}.  Both methods have been applied successfully, the density method in \cite{tdf2,GHR}, the
density matrix method more extensively in 
\cite{tdf6,GHR,BA1,BA2,HG1,HG2,WSF1,WSF2}.  The mathematical
equivalence of the two formalisms derived from TDDFT has been established in Ref.~\cite{GHR}, so that it is only
a matter of numerical convenience which one is used in practice.
To argue the utility of the formalism developed in this paper,
we shall show that in a suitable approximation it yields the eigenvalue equation of the density-matrix CKS scheme.  It also
permits the calculation of electromagnetic transition rates
between the excited states studied and the ground state without the explicit use 
of wave functions.

We also call attention to several recent studies of the
excited state problem that involve extensions of the variationally based
KS theory to individual excited states \cite{AG,LG}.  For these methods, as well,
successful applications have been made to simple systems.  
Improved exchange and correlation kernels necessary for all these methods and
a connection with many-body perturbation theory are discussed in \cite{GS},
whereas in \cite{TH} an exchange-correlation potential is suggested
to provide more accurate continuum KS orbitals needed for excited state
and polarizability calculations.

In this paper, we return
to a study of the trace variational principle \cite{kk1,kk2,kk3}.  
Instead of considering the average energy as
a functional of the average density, however, we argue for the introduction
of a matrix array of densities, i.\ e., all matrix elements of the density
operator among all states of a chosen ensemble, and for an investigation
of the average energy as a functional of this matrix array.  In Sec.\ II
we present arguments to indicate how the Hohenberg-Kohn (HK) analysis can be extended to
this case yielding a variational equation for the matrix.  We subsequently
(Sec.\ III) generalize the KS analysis, deriving a matrix Kohn-Sham equation
(MKS), that contains not only the expected ingredient, a matrix effective
potential, but also a matrix of Lagrange multipliers arising from number
conservation in each state of the chosen subset; this matrix can be
diagonalized, but not otherwise transformed away.  By combining
solutions of the MKS equations, we can construct the density array.

As an application of this theory, we study, in Sec.\ IV,
the MKS equations in the weak-coupling limit, which
is the equivalent, in our approach, to linear response
theory.  In this limit,
we include only the ground state and excited states characterized 
(largely) by one 'quasiparticle-quasihole' excitations of the ground-state.  
Higher excited states are incorporated via simple assumptions concerning their
properties.   The
major result of this analysis is an eigenvalue equation for the
aforementioned Lagrange multipliers (relative to their ground-state
value) that has the same form as the eigenvalue equation of the 
density-matrix based CKS.  
Assuming that the ground-state KS problem has been solved,
the major unknown ingredient in these equations, an exchange-correlation
interaction, can be identified with the corresponding quantity
utilized in current applications, which in practice require  an adiabatic approximation.

Comparison of our result with the existing formalism establishes the identity of the eigenvalues of our equation with excitation
energies of the system.  These energies can be calculated as well
from a difference of adjacent averages of the Hamiltonian. By
this means, in Sec.~V, we establish within the framework of our formalism the physical meaning of the eigenvalues.  In a concluding section, we summarize our considerations.

\section{Hohenberg-Kohn arguments}

The Hamiltonian is written as 
\begin{equation}
\hat{H} = \hat{T} + \hat{V} + \hat{W} +\hat{Y},  \label{ham}
\end{equation}
the sum of the kinetic energy, the electrostatic interaction of the 
electrons with the nucleus, the Coulomb repulsion of the electrons,
and an additional fictitious external source term that will be set to zero
for actual calculations.
The various terms have the forms (${\bf x}$ stands
for the space-spin pair $({\bf r},s)$), in atomic units,
\begin{eqnarray}
\hat{T} &=& \int d{\bf x} \hat{\psi}^{\dag}({\bf x})(-\frac{1}{2}\nabla^2)
\hat{\psi}({\bf x})  \nonumber \\
   &=&  \int \hat{\psi}^{\dag}\tau\hat{\psi},   \label{def1} \\
\hat{V} &=& \int d{\bf x}\hat{\psi}^{\dag}({\bf x})\hat{\psi}({\bf x})
v({\bf r}),  \label{def2} \\
\hat{W} &=& \int d{\bf x}d{\bf x'} \frac{1}{|{\bf r-r'}|}
  \hat{\psi}^{\dag}({\bf x})\hat{\psi}^{\dag}({\bf x'})
  \hat{\psi}({\bf x'})\hat{\psi}({\bf x}),   \label{def3} \\
\hat{Y} &=& \int d{\bf x}d{\bf x'} y({\bf x,x'})\hat{\eta}({\bf x,x'}),
                \label{def4}  \\
\hat{\eta} &=& \hat{\psi}^{\dag}({\bf x})\hat{\psi}({\bf x})
                 \hat{\psi}^{\dag}({\bf x'})\hat{\psi}({\bf x'}).\label{def5}
\end{eqnarray}
The interaction term $\hat{Y}$ is a combination of one and two body forces.
It serves as a device to establish the dependence of the theory on the
extended density defined in Eq.~(\ref{defn}), but is set equal to zero thereafter.  This
procedure is akin to the use of an external magnetic field in standard density 
functional theory in order to exhibit the possible dependence on spin
polarization.
For the traces of these operators over the ensembles introduced below,
we use the same symbols without hats.

In the following we shall base our arguments on the variational principle
for the trace of the Hamiltonian over the lowest $M$ eigenstates of a many body
system \cite{tr1,tr2,tr3,kk1,kk2,kk3}.  We consider the case where the
$(M+1)${\it st} state has a
higher energy than the $M${\it th} state, although this criterion is not
absolutely necessary.

Let
\begin{equation}
{\cal S} = \{|I\rangle \}   \label{set}
\end{equation}
be the space of states included ($I=1...M$).
For any operator $\hat{O}$, we define the restricted trace
\begin{equation}
O^{(M)} = \sum_{I=1}^{M} \langle I|\hat{O}|I\rangle,   \label{def10}
\end{equation}
where it is convenient in the further development not to divide by $M$.

We then consider a set of propositions formulated in imitation of the Hohenberg-Kohn (HK) theorem \cite{KD}:

(i) Every choice of a function $y({\bf x,x'})$ in (\ref{def4}) determines an $M$-dimensional 
space ${\cal S}$ through the solution of the Schr{\"o}dinger equation.  Two different functions $y\neq y'$ will yield
different ${\cal S}\neq {\cal S}'$, provided $\hat{Y}[y]$
and $\hat{Y}[y']$ differ by more than a diagonal matrix in the 
space ${\cal S}$.  To see this, suppose that $\hat{Y}[y]$ and
$\hat{Y}[y']$ yield the same space ${\cal S}$.  From the Schr{\"o}dinger equations for the state $|I\rangle$, we obtain by
subtraction that
\begin{equation}
\{\hat{Y}[y]-\hat{Y}[y']\}|I\rangle=(E_I-E_I^\prime)|I\rangle,
\label{equivY}
\end{equation}
where the $E's$ are the corresponding eigenvalues.  Thus, with
$I' \neq I$, 
\begin{equation}
\langle I'|\hat{Y}[y]|I\rangle=\langle I'|\hat{Y}[y']|I\rangle,
\label{ndY}
\end{equation}
and
\begin{equation}
\hat{Y}[y]-\hat{Y}[y']=\sum_{I=1}^M (E_I-E_I^\prime)|I\rangle\langle I|.
\label{dY}
\end{equation}
Potentials that satisfy this relation will be considered equivalent.

(ii) ${\cal S}$ determines the correlation function $\eta({\bf x,x'})=\sum\langle I|\hat{\eta}({\bf x,x'})|I\rangle$.
This relationship is single-valued and invertible.  This can be proved
by an adaptation of the standard HK argument, as we now show.  Suppose that
\begin{equation}
{\cal S} \rightarrow \eta,\;\;  {\cal  S}'\neq {\cal S} \rightarrow \eta'.
\end{equation}
It follows that $\eta\neq\eta'$.  We prove this by using the
trace variational principle (valid by construction of the sets, as indicated above)
 to establish two inequalities,
\begin{eqnarray}
H_{{\cal S}}[y] &<& H_{{\cal S}'}[y']+\int(y-y')\eta',  \label{in1}  \\
H_{{\cal S}'}[y'] &<& H_{{\cal S}}[y] +\int(y'-y)\eta.  \label{in2}
\end{eqnarray}
Here, for example, $H_{{\cal S}}[y]$ is the ensemble average of $\hat{H}$
over the set ${\cal S}$, where it is further emphasized that this average
is a functional of $y$.  Adding (\ref{in1}) and (\ref{in2}) and
assuming that $\eta=\eta'$, we obtain the usual contradiction
\begin{equation}
H_{{\cal S}}[y] + H_{{\cal S'}}[y']<H_{{\cal S}'}[y'] + H_{{\cal S}}[y].
\end{equation}
Thus ${\cal S}$ is a single-valued functional of $\eta$.

Considering $H$ to be a functional of $\eta$, we write the 
variational principle in the form 
\begin{equation}
\delta H =\int\frac{\delta H}{\delta\eta}\delta \eta =0. \label{var1}
\end{equation}
We shall not attempt, however, to implement the variational principle
in this version.  Instead, using completeness, we introduce the relation
\begin{eqnarray}
\eta({\bf x,x'}) &=&\sum_{I=1}^M\sum_{I'=1}^\infty
\langle I|\hat{\psi}^{\dag}({\bf x})\hat{\psi}({\bf x})|I'\rangle \nonumber\\
&&\times \langle I'|\hat{\psi}^{\dag}({\bf x'})\hat{\psi}({\bf x'})|I\rangle.
\label{compl}
\end{eqnarray}

As long as $M$ is finite, this sum is asymmetric in the two sets of indices.  However our aim
is to utilize as variational parameters the quantities
\begin{equation}
n({\bf x})_{I'I} =\langle I|\hat{\psi}^{\dag}({\bf x})
                  \hat{\psi}({\bf x})|I'\rangle \;\;\; 
\{I,I'\epsilon{\cal S}\} ,     \label{defn}
\end{equation}
which constitute the elements of an $M\times M$ square matrix
${\bf n}$.  There are two arguments that can be put forward to satisfy ourselves that we can consider the correlation function $\eta({\bf x,x'})$ to be a functional of this more restricted set of matrix elements.
The first and more satisfactory one at this stage is to let the number $M$ of states included in the trace become large enough so that either pointwise
or in the norm the fractional contribution from pieces connecting $I\leq M$ to
$I>M$ is negligible.  The second is to ask for the temporary indulgence of the reader, by guaranteeing to show that in any {\it application}, these matrix elements can either be ignored or expressed explicitly in terms of those of 
{\bf n}.  We call this a closure approximation.  When we
turn to the application developed in Secs.~IV and V, we shall see,
 that the accuracy of results obtained there depend on a choice of a
closure approximation that is historically tied to linear response theory
and the associated random phase approximation, namely the approximate characterization of the simplest excited states as boson excitations.

In consequence of the arguments given above,
we replace the variational principle (\ref{var1}) by the form
\begin{equation}
\delta H = \int\frac{\delta H}{\delta {\bf n}}\delta {\bf n}.  \label{veta}
\end{equation}
From Eq.\ (\ref{veta}) we can derive a variational equation
by imposing the number conservation constraints.  If $N$ is the number of
electrons, we have
\begin{equation}
\int d{\bf x} n({\bf x})_{II'} = N\delta_{II'}.
\end{equation}
Introducing a set of Lagrange multipliers $\mu_{II'}$, we now write
\begin{equation}
\delta H -\mu_{II'}\int\delta n({\bf x})_{I'I} =0,   \label{var3}
\end{equation}
and conclude that
\begin{equation}
\frac{\delta H}{\delta n({\bf x})}_{I'I} = \mu_{II'} \;. \label{gtf}
\end{equation}

In concluding this section, we emphasize once more that our purpose in introducing the unusual "external" field $\hat{Y}$ was to provide
a pathway leading to the formulation, 
Eqs.~(\ref{veta})-(\ref{gtf}), which form the basis for the 
further development.  These relations also apply to systems with
$\hat{Y}=0$, which are the {\em only} systems that interest us henceforth.

\section{Generalized Kohn-Sham scheme}

$n({\bf x})_{II'}$ is the limit ${\bf x}\rightarrow {\bf x'}$ of the
off-diagonal one-body density matrix
\begin{equation}
\rho({\bf x}I|{\bf x'}I') =\langle I'|\hat{\psi}^{\dag}({\bf x'})
                          \hat{\psi}({\bf x})|I\rangle.  \label{denm}
\end{equation}
Since $\rho$ is Hermitian and (as we shall show below) positive
semi-definite in the direct product space labeled
by $({\bf x},I)$, it can be brought to diagonal
form, a move that generalizes the concept of natural orbitals.  We may thus write
\begin{eqnarray}
\rho({\bf x}I|{\bf x'}I') &=& \sum_J \lambda_J\Phi_J({\bf x}I)
                              \Phi_J^{\ast}({\bf x'}I'),  \label{nat1} \\
\lambda_J &\geq& 0,   \label{nat2} \\
\sum_{I=1}^M\int d{\bf x}\,\Phi_J^{\ast}({\bf x}I)\Phi_{J'}({\bf x}I)
                 &=& \delta_{JJ'},  \label{nat3} \\
\int d{\bf x}\,\rho({\bf x}I|{\bf x}I') &=& N\delta_{II'}.  \label{nat4}
\end{eqnarray}
Here Eqs.~(\ref{nat1}) and (\ref{nat2}) define the eigenfunctions and eigenvalues
of the generalized density matrix, (\ref{nat3}) expresses the property
that the $\Phi_J({\bf x}I)$ are unit eigenvectors in the space labeled jointly
by the single-particle coordinates and the eigenvalues of the states in the
set ${\cal S}$, and (\ref{nat4}) expresses number conservation.  It
follows from these equations that
\begin{equation}
\sum_{I=1}^M\int d{\bf x}\,\rho({\bf x}I|{\bf x}I) = \sum_J\lambda_J =NM.\label{nat5}
\end{equation}

Before continuing the development, we interject the proof
that $\rho({\bf x}I|{\bf x'}I')$ is positive semi-definite.
Toward this end we compare Eq.~(\ref{nat1}) with the form
that follows directly from its definition
\begin{equation}
\rho({\bf x}I|{\bf x'}I')=\sum_K\langle I'|\hat{\psi}^{\dag}
({\bf x'})|K\rangle\langle K|\hat{\psi}({\bf x})|I\rangle,
\label{psd1}
\end{equation}
where $K$ is a complete set of intermediate states.  Invoking the
orthonormality relations (\ref{nat3}), we thus conclude that
for eigenvalues $\lambda_J\neq 0$,
\begin{equation}
\lambda_J =\sum_K|\sum_{I=1}^M\int d{\bf x} \Phi_J^{\ast}({\bf x}, I)
\langle K|\hat{\psi}({\bf x})|I\rangle|^2 > 0. \label{psd2}
\end{equation}

In imitation of ground-state KS theory, we introduce a mapping
from the off-diagonal density
to a quasi-independent-particle off-diagonal density,
\begin{eqnarray}
n({\bf x})_{II'} &\rightarrow& n^s({\bf x})_{II'},  \label{ks1} \\
n^s({\bf x})_{II'} &=& \sum_J \varphi_J({\bf x}I)
                              \varphi_J^{\ast}({\bf x}I'),  \label{ks2} \\
\sum_{I=1}^M\int d{\bf x}\,\varphi_J^{\ast}({\bf x}I)\varphi_{J'}({\bf x}I)
                 &=& \delta_{JJ'},  \label{ks3} \\
\int d{\bf x} \, n^s({\bf x})_{II'} &=& N\delta_{II'}.  \label{ks4}
\end{eqnarray}
Though we use the same symbol $J$ to label orbitals as for the case of
natural orbitals, here the similarity stops.  For the latter, $J$ is, in
principle, an unbounded set.  For the present alternative, the set labeled by
$J$ is strictly a finite set as determined by the sum (cf. (\ref{nat5})),
\begin{equation}
\sum_J 1 = NM.     \label{ks5}
\end{equation}
Though the labels $I$ have the same meaning in both the original space and in the mapped space, the fact that the labels $J$ do
not, leaves a vagueness at present about the significance of this quantum number.  We shall see, however, that once we turn to applications in Sec.~IV, there will be no problem in providing
a precise characterization of this quantum label. 

We next show how the variational principle may be used to obtain
equations for the orbitals $\varphi_J$ so that in fact the matrices
${\bf n}$ and ${\bf n}^s$ are equal. We utilize the variational
principle in the form
\begin{equation}
\sum_{JI}\int d{\bf x}\frac{\delta H}{\delta \varphi_J^{\ast}({\bf x}I)}
\delta\varphi_J^{\ast}({\bf x}I) +\;\;{\rm c.c.} =0\,.   \label{var4}
\end{equation}
Imitating the
procedure for the ground-state theory, we decompose the trace of the Hamiltonian, 
\begin{eqnarray}
H &=& T^s +(V+W+ T-T^s),  \label{KS1}    \\
T^s &=& \sum_J \int \varphi_J^{\ast}t\varphi_J.  \label{KS2}
\end{eqnarray}
Enforcing the equality of ${\bf n}$ and ${\bf n}^s$, we define an
effective single-particle potential matrix,
\begin{eqnarray}
v^s({\bf x})_{II'} &=& \frac{\delta}{\delta n({\bf x})_{I'I}}
                       (V+W+T-T^s)  ,           \label{KS3} \\
                   &=& \frac{\delta}{\delta n^s({\bf x})_{I'I}}
                        (V+W+T-T^s).      \label{KS4}
\end{eqnarray}
The discussion of the decomposition of this matrix 
single-particle operator into constituent parts of interest will be taken up in Sec.~IV.

With the help of Eqs.~(\ref{KS1}-\ref{KS4}), 
we derive from the variational
principle (\ref{var4}) the conditions
\begin{eqnarray}
&&\sum_{JII'}\int\, d{\bf x}\delta\varphi_J^{\ast}({\bf x}I)[\tau\delta_{II'}
 + v^s({\bf x})_{II'}]\varphi_J({\bf x}I') + \;\;{\rm c.c.} =0. \label{sp1}
\end{eqnarray}   
To derive generalized single-particle equations of motion from the
variational principle, we add the constraint conditions  
\begin{equation}
-\sum_{JII'}\int\, d{\bf x}\delta\varphi_J^{\ast}({\bf x}I)[\epsilon_J\delta_{II'}
+\nu({\bf x})_{II'}]\varphi_J({\bf x}I') \;\;+{\rm c.c.}=0\;\;.  \label{sp2}
\end{equation}
Here $\epsilon_J$ is the Lagrange multiplier for the normalization condition
contained as part of (\ref{ks3}). (As usual, the orthogonality condition
need not be imposed, since it will be automatically satisfied by the
solutions of the emerging equations.)
The unfamiliar term containing the Lagrange multiplier matrix
$\nu({\bf x})_{II'}$ has the form of an additional potential matrix,
whose purpose is to enforce the condition \cite{no} that ${\bf n}={\bf n}^s$.  We
shall study this quantity further below.
Combining Eqs.~(\ref{sp1}) and (\ref{sp2}), we derive (together with it complex
conjugate) the generalized single-particle equation
\begin{equation}
\epsilon_J\varphi_J({\bf x}I)=\sum_{I'}[\tau \delta_{II'} +v^s({\bf x})_{II'}
-\nu^s({\bf x})_{II'}]\varphi_J({\bf x}I').   \label{eom1}
\end{equation}

At this juncture it is appropriate to wonder how (\ref{eom1}) is related to the density and density-matrix forms of CKS theory. 
We cannot
expect a general connection, since the latter describes the consequences
of the application of a time-dependent external field, whereas in the 
theory under development, the ``time dependence'' is a purely internal
matter expressed by an off-diagonal array of densities and effective
potentials.  Nevertheless, a connection between the two formalisms 
will be made  
for the application studied in Sec.\ IV, the so-called weak-coupling limit.  In effect, it will be shown that this limit
contains the same physics as the combination of TDDFT and linear response theory, and thus solves, in part, the task of establishing the "usefulness" of our approach.

We conclude the present section by showing that (cf.~Eq.~(\ref{gtf}))
\begin{equation}
\nu({\bf x})_{II'} =\mu_{II'},  \label{lm1}
\end{equation}
up to a global additive constant.
It is thus a non-trivial matrix and cannot be absorbed into the eigenvalues
$\epsilon_J$.
To prove (\ref{lm1}), we can work backwards from the sum of (\ref{sp1})
and (\ref{sp2}) to the equation
\begin{eqnarray}
0 &=& \sum_{II'}\int\,d{\bf x}\left[\frac{\delta H}{\delta n^s({\bf x})_{II'}}
-\nu({\bf x})_{II'}\right]\delta n^s({\bf x})_{I'I}  \label{lm2} \\
 &=& \sum_{II'}\int\,d{\bf x}\left[\frac{\delta H}{\delta n({\bf x})_{II'}}
-\nu({\bf x})_{II'}\right]\delta n({\bf x})_{I'I}  \label{lm3} \\
&=& \sum_{II'}\int\,d{\bf x}\left[\frac{\delta H}{\delta n({\bf x})_{II'}}
-\mu_{II'}\right]\delta n({\bf x})_{I'I}.  \label{lm4}
\end{eqnarray}
In passing from (\ref{lm2}) to (\ref{lm3}), we have used the equality
${\bf n^s}={\bf n}$.  In writing (\ref{lm4}), we have repeated (\ref{var3}).
Comparing  (\ref{lm3}) with (\ref{lm4}), we arrive at (\ref{lm1}), again
up to an additive constant. 

\section{Application to the weak coupling limit}

In the course of this section, we shall transform and approximate
Eq.~(\ref{eom1}), leading to an eigenvalue equation that will determine
off-diagonal elements of the matrix ${\bf n}$.
We shall do so in an approximation, the weak-coupling approximation,
that is equivalent to a linear response approach.  Assuming that the matrix ${\bf \mu}$ can be chosen diagonal (see immediately below),
the eigenvalues are the quantities
\begin{equation}
\omega_I  =\mu_{II} -\mu_{00}.  \label{omeg1}
\end{equation}
The proof that the matrix ${\bf \mu}$ can be chosen diagonal goes as
follows: Though we trace over a set of states labeled $I$ and originally
identified as eigenstates of the reference system, the entire formalism
is invariant under a unitary transformation within the included space.
Such a transformation can be chosen to diagonalize ${\bf \mu}$ if it is not already diagonal.
The relation of the quantities in Eq.\ (\ref{omeg1}) to the
excitation energies of the system is not immediately apparent. The main result of this section suggests that they are equal.
The proof that they are is given in Sec.~V.

Though the derivation of the main result of this section, the eigenvalue
equation, can be carried out directly from the generalized KS equation,
we present the discussion in a form that makes more
immediate contact with the density functional form of the theory.  
The first step, which is completely general, is to transform Eq.~(\ref{eom1})
into an equation for the matrix $n^s_{II'}({\bf x,x'})$.  First rewrite
Eq.~(\ref{eom1}), remembering Eq.~(\ref{omeg1}), as
\begin{eqnarray}
\epsilon_J \varphi_J({\bf x}I) &=& \sum_{I'}(h_{II'}^s({\bf x})-\omega_I\delta_{II'})\varphi_J({\bf x} I'),  \label{eom10} 
\end{eqnarray}
Recalling the definition 
\begin{equation}
n^s_{II'}({\bf xx'}) =\sum_J \varphi_J({\bf x}I)\varphi_J^{\ast}({\bf x}'I'),
\label{def100}
\end{equation}
we can form from Eq.~(\ref{eom10}) and its complex conjugate two equivalent
but distinct values of the sum
$\sum_J \epsilon_J \varphi_J({\bf x}I)\varphi_J^{\ast}({\bf x}'I')$. 
The difference of these forms yields the generalized density-matrix equation (using summation convention)
\begin{equation}
n^s_{II'}({\bf xx}')(\omega_{I'}-\omega_I) = \sum_{I''}[n^s_{II''}({\bf xx}')
h^s_{I''I'}({\bf x}') - h^s_{II''}({\bf x})n^s_{I''I'}({\bf xx'})],
\label{dm10}
\end{equation}
that will provide the starting point for our further considerations.

We note that by introducing time-dependent
matrix elements
\begin{equation}
O_{II'}(t) \equiv O_{II'}\exp[-i(\omega_I -\omega_{I'})t], \label{tdel}
\end{equation}
where $O$ takes on the values $n^s$ and $h^s$, Eq.~(\ref{dm10}), may be written
in the form
\begin{equation}
-i\frac{d}{dt}{\bf n}^s(t) =[{\bf n}^s(t),{\bf h}^s(t)].  \label{dm11}
\end{equation}
This resembles the fundamental equation of TDDFT, in density matrix form,
except that the bold-face type reminds us that we are dealing with 
quantum-mechanical operators rather than c-numbers.  This can be converted
into a form of TDDFT, however, by assuming the existence of a wave packet
$|\Psi\rangle$ that is a linear combination of the ground state and excited
states of interest, for which we can also replace the average of the products
that appear in the commutator by the product of the averages.  

In the following we concentrate on the study of the weak coupling approximation
to Eq.~(\ref{dm10}).
In this approximation one imagines that the set of states (\ref{set}) can be divided 
into the ground-state $I = 0$ and states characterized as dominant $\nu$ quasiparticle-
$\nu$ quasihole excitations (denoted by $I_{\nu}$) with respect to the ground-state.
On the basis of this picture one may introduce  a set of assumptions concerning
relative orders of magnitude of certain matrix elements, whose
validity is obvious in the limit of vanishing two-particle interaction,
\begin{eqnarray}
|n^s_{00}| &>>& |n^s_{0I_1}| >> |n^s_{0I_2}|>> ...,  \label{in11}\\
|n^s_{I_1 I_1}| & \approx& |n^s_{00}|,   \label{in12} \\
|n^s_{I_1 I_1'}| &\approx& |n^s_{0I_2}| \;\;\;{\rm if}\;
I_1\neq I_1'.   \label{in13}
\end{eqnarray}
We shall consider diagonal elements to be of zero order, elements connecting
states $I_\nu$ to $I_{\nu +p}$ to be of $p$th order.

Considering assumption (\ref{in12}) first, it asserts that
 for $I$ belonging
to the first few levels of the hierarchy, if $N$, the number of particles is not too small,
in lowest approximation matrix elements diagonal in $I$ are equal to
their value for $I=0$.  It is easiest to see this for the density itself,
since the wave functions of the excited states differ from those of the
ground state by at most a few particles out of $N$.  That this property of matrix elements diagonal in $I$ follows for 
quantities other than the density itself is a consequence of their relation to the density, as will
be seen from further study below. We shall consider all diagonal matrix
elements to be zero order quantities. 
A further assumption, in terms of this scale, is that matrix elements
in which $I$ and $I'$ belong to adjacent levels in the hierarchy are,
on the average, of order $(1/\sqrt{N})$ compared to zero order quantities.
For the sorting of our equations, we also need the assumption that matrix
elements in which $I,I'$ differ by two levels or refer to two different
states of the same level are second order quantities, i.\ e., of the order
of the product of first order quantities.  Of course, it has to be verified
{\it a posteriori} that the solutions found are in accord with these
statements.

Our aim is to apply these assumptions to choose those matrix elements of Eq.~(\ref{dm10}) that characterize the state $0$
and the states $I_1$. To carry out this program, we must look
more closely into the structure of the effective interaction
${\bf v}^s$.  First we rewrite the trace of the Hamiltonian
in the form
\begin{eqnarray}
H &=& T^s +V + W^c +H^{xc},      \label{xc1} \\
W^c &=& \frac{1}{2}\int d{\bf x}d{\bf x'}\sum_{II'}n^s_{II'}({\bf x})\frac{1}{|{\bf x} -{\bf x}'|}n^s_{I'I}({\bf x'}),  \label{xc2}
\end{eqnarray}
which defines $H^{xc}$.  It follows that
\begin{eqnarray}
v^s_{II'}({\bf x}) &=& \frac{\delta}{\delta n^s_{I'I}({\bf x})}
(V +W^c +H^{xc})  \\
 &=& v({\bf x})\delta_{II'} +v^c_{II'}({\bf x})
+v^{xc}_{II'}({\bf x}), \label{vxc1}  \\
v^c_{II'}({\bf x}) &=& \int d{\bf x'}\frac{1}{|{\bf x}-{\bf x}'|}n^s_{II'}
({\bf x'}).  \label{vxc2} 
\end {eqnarray}
The main reason for exhibiting these formulae is to recognize,
as we shall see in more detail below, that the off-diagonal elements of ${\bf h}$ are at least linear in the corresponding
off-diagonal elements of ${\bf n}^s$.  This is obvious from 
Eq.~(\ref{vxc2}) for the Coulomb contribution and will be argued
more closely later for ${\bf v}^{xc}$. Thus we may safely
assume that that the matrix elements of ${\bf h}$ are the same
order of magnitude as the corresponding matrix elements of 
${\bf n}^s$.

Turning then to our major task, which is to study the matrix elements of Eq.~(\ref{dm10}), we consider first the ground or 00 element.  Neglecting terms of second order and higher, we find
\begin{equation}
 [n^s_{00}({\bf xx}')h^s_{00}({\bf x'}) - h^s_{00}({\bf x})
n^s_{00}({\bf xx'})] =0.    \label{ksden0}
\end{equation}
It is consistent with our approximations to identify $n^s_{00}$ 
(in leading approximation only) with the ground
state density of KS theory and $h^s_{00}({\bf x})$ with the KS single-particle Hamiltonian.
Equation (\ref{ksden0}) is thus the KS equation in density matrix form and determines a complete set 
of quasiparticle orbitals $\varphi_a({\bf x})$, where $a=h$ will refer
to the quasi-orbitals occupied in the ground-state and $a=p$ to those unoccupied. 

Consider next the first-order matrix element $01$.  Retaining only first-order
contributions (leading corrections are third order), we may write
\begin{eqnarray}
\omega_1 n^s_{01}({\bf xx}')&=&[n^s_{00}({\bf xx}')h^s_{01}({\bf x'})+
n^s_{01}({\bf xx}')h^s_{11}({\bf x'})\nonumber \\
&&-h^s_{00}({\bf x})n^s_{01}({\bf
xx'}) -h^s_{01}({\bf x})n^s_{11}({\bf xx'})] . \label{ksden1}
\end{eqnarray}
As a first step in the evaluation of this equation, we may, according to 
Eq.~(\ref{in12}), set the $n_{11}$ matrix elements equal to the $n_{00}$ ones.  We also
drop the subscripts $00$ understanding these according to the previous
identification to be the standard KS quantities.  If we can exhibit 
$h^s_{01}$ as an (approximate) linear functional of $n^s_{01}$, Eq.~(\ref{ksden1})
will have the form of a linear eigenvalue problem.  First we have (the matrix
elements in question are local functions of ${\bf x}$)
\begin{eqnarray}
h^s_{01}({\bf x}) &=& v^c_{01}({\bf x}) +v^{xc}_{01}({\bf x}), \label{hxc100} \\
v^c_{01}({\bf x}) &=& \int d{\bf x'}\frac{1}{|{\bf x}-{\bf x}'|}n^s_{01}({\bf x}'). \label{hxc101}
\end{eqnarray}
We see that $v^c$ is, by definition, already of the desired form.

We turn then to $v^{xc}$.  Our approach to this quantity is to 
revert
to the study of $H^{xc}$, defined in Eq.~(\ref{xc1}), 
which we consider, in line with assumptions previously made, 
a functional of $n_{00}\approx n$, of $n^s_{01}$, and of $n^s_{10}$, the latter two considered as small quantities.  (It is also a functional of the other
off-diagonal elements, $n^s_{01'}$ and $n^s_{1'0}$, where $1'$ refers to any of
the other states at level one of the hierarchy of states.  It is simply that
this dependence does not enter into the current discussion.)
We then expand $H^{xc}$ as a functional
Taylor series in these quantities,
\begin{eqnarray}
H^{xc} &=& H^{xc}|_0 +\int d{\bf x}\frac{\delta H^{xc}}{\delta n^s_{10}({\bf x})}|_0 n^s_{10}({\bf x}) +\int d{\bf x}\frac{\delta H^{xc}}{\delta n^s_{01}({\bf x})}|_0 n^s_{01}({\bf x}) \nonumber \\
&&+\int d{\bf x}d{\bf x'}\frac{1}{2}\frac{\delta^2 H^{xc}}{\delta n^s_{10}({\bf x})\delta n^s_{10}({\bf x'})}|_0 n^s_{10}({\bf x})n^s_{10}({\bf x'})    
+\int d{\bf x}d{\bf x'}\frac{\delta^2 H^{xc}}{\delta n^s_{10}({\bf x})\delta n^s_{01}
({\bf x'})}|_0 n^s_{10}({\bf x})n^s_{01}({\bf x'}) \nonumber \\
&&+\int d{\bf x}d{\bf x'}\frac{1}{2}\frac{\delta^2 H^{xc}}{\delta n^s_{01}({\bf x})
\delta n^s_{01}({\bf x'})}|_0 n^s_{01}({\bf x})n^s_{01}({\bf x'})
+... \;\;.   \label{fts}
\end{eqnarray}

Strictly, the quantity $H^{xc}|_0$ and its functional derivatives still depend on $n_{11}$ as well as $n_{00}$.  It suffices to ignore the difference of the
two quantities in the present discussion, but we shall have to remember and include the difference in the arguments of Sec.\ V.
We note further that only the first and fourth of the terms shown explicitly in this equation are non-vanishing.  Recall that $H^{xc}$ is a trace
and therefore invariant under a unitary transformation in the space of states $I$.  Its dependence on the matrix ${\bf n}$
must also be in the form of traces over these indices.  As we can see
on the example of the Coulomb interaction, this dependence is more general than traces of products of ${\bf n}$ at the same point, but in any event it follows that for every factor of 
$n^s_{10}$ at some spatial point, there must be a factor of
$n^s_{01}$, at a generally different point.  The simplification described above follows.  We thus compute to first order
\begin{eqnarray}
v^{xc}_{01}({\bf x}) &=& \int d{\bf x'}\frac{\delta^2 H^{xc}}{\delta n^s_{10}
({\bf x})\delta n^s_{01}({\bf x'})}|_0 n^s_{01}({\bf x'}) \nonumber \\
&\equiv& \int d{\bf x'}f_{10,10}(|{\bf x} -{\bf x'}|,n)n^s_{01}({\bf x'}) \nonumber \\
&\approx& \int d{\bf x'}f(|{\bf x}-{\bf x'}|,n)n^s_{01}({\bf x'}).  \label{linxc}
\end{eqnarray}
In passing from the second to the third line of this equation, i.\ e., in ignoring
the state-dependence of $f$, we are making an approximation equivalent to the 
adiabatic approximation widely used in TDDFT.
With the definition (the dependence on $n$ being understood)
\begin{equation}
f^{eff}(|{\bf x} - {\bf x'}|)= \frac{1}{|{\bf x} -{\bf x'}|}
                                +f(|{\bf x}-{\bf x'}|),  \label{defef}
\end{equation}
Eq.~(\ref{ksden1}) may be rewritten as (using $h^s({\bf xx''})=\delta({\bf x}-{\bf x'}) h^s({\bf x})$
for convenience)
\begin{eqnarray}
\omega_1 n^s_{01}({\bf xx'})&=& \int\, d{\bf x''}[n^s({\bf xx'})f^{eff}(|{\bf x'}-{\bf x''}|)n^s_{01}({\bf x''})
+n^s_{01}({\bf xx''})h^s({\bf x''x'})   \nonumber \\
&& -h^s({\bf xx''})n^s_{01}({\bf x''x'})
   -n^s({\bf xx'})f^{eff}(|{\bf x}-{\bf x''}|)n^s_{01}({\bf x''})]. 
\label{rpa0}
\end{eqnarray}.

The final task with respect to this equation is to convert it into a standard
RPA form.  Toward this end we reexpress the matrices $n^s$ and $n^s_{01}$
in terms of the KS single-particle functions, $\varphi_a({\bf x})$, satisfying the KS equation
\begin{eqnarray}
\int\, d{\bf x'}h^s({\bf xx'})\varphi_a({\bf x'}) &=& \epsilon_a\varphi_a({\bf x}).
\label{kseq} 
\end{eqnarray}
First of all we have the familiar equation
\begin{equation}
n^s({\bf xx'}) = \sum_h \varphi_h({\bf x})\varphi_h({\bf x'}).  \label{gsden}
\end{equation}
Next we must evaluate the sum
\begin{equation}
n^s_{01}({\bf xx'}) = \sum_J\varphi_J({\bf x}0)\varphi^{\ast}_J
({\bf x'}1).   \label{odden1}
\end{equation}
Here we must introduce assumptions concerning which values 
of $J$ contribute to the required order.  In the space of the 
eigenstates of the fully interacting system, we are concerned with the ground
state and with states that are quasiparticle-hole (qph) excitations of this state.  When we remove one particle from (create a hole $h$ in) such a state, we expect to encounter states that can be characterized as either $0h$ or $1h$, and these are the values of $J$
that we assign in the sum (\ref{odden1}).  If we consistently use the 
approximations $\varphi_{0h}(0)\approx \varphi_{1h}(1)\approx \varphi_h$,
the weak-coupling value of Eq.~(\ref{odden1}) becomes
\begin{equation}
n^s_{01}({\bf xx'})=\sum_h [\varphi_h({\bf x})\varphi^{\ast}_{0h}({\bf x'}1)
+\varphi_{1h}({\bf x}0)\varphi^{\ast}_h({\bf x'})].   \label{odden2}
\end{equation}

The final form for this quantity is achieved by expanding the first-order amplitudes in terms of KS modes,
\begin{eqnarray}
\varphi_{0h}(1) &=& \sum_p\varphi_p X_{ph},    \label{rpax} \\
\varphi_{1h}(0) &=& \sum_p\varphi_p Y^{\ast}_{ph}.    \label{rpay}
\end{eqnarray}
The restriction of the sums on the right-hand sides of these equations
is also consistent with the weak-coupling picture painted above.  Strictly
the amplitudes $X,Y$ should carry superscripts $1$, identifying the eigenstate
to which they refer, but we shall suppress these except when required for clarity,
as in Sec.\ V.  Finally then,
\begin{equation}
n^s_{01}({\bf xx'})=\sum_{p,h}[\varphi_h({\bf x})\varphi^{\ast}_p({\bf x'})X_{ph}^{\ast}
+\varphi_p^{\ast}({\bf x})\varphi_h({\bf x'})Y^{\ast}_{ph}].  \label{odden3}
\end{equation}

Introducing Eqs.~(\ref{gsden}) and (\ref{odden3}) into Eq.~(\ref{rpa0}),
we can project out equations for $X^{\ast}_{ph}$ 
and $Y^{\ast}_{ph}$.  We quote the complex conjugate of these equations:
\begin{eqnarray}
(\epsilon_h -\epsilon_p +\omega_1)X_{ph} &=& (f^{eff})_{ph'hp'}
X_{p'h'}  +(f^{eff})_{pp'hh'}Y_{p'h'},  \label{RPA1} \\
(\epsilon_h -\epsilon_p -\omega_1)Y_{ph} &=&
(f^{eff})_{hp'ph'}Y_{p'h'}+(f^{eff})_{hh'pp'}X_{p'h'},  \label{RPA2}  \\
(f^{eff})_{abcd}&=& \int d{\bf x}d{\bf x'}\varphi_a^{\ast}({\bf x})\varphi^{\ast}_b({\bf x'})
f^{eff}(|{\bf x}-{\bf x'}|)\varphi_c({\bf x})\varphi_d({\bf x'}). \label{RPA3}
\end{eqnarray}

The equations found are of the same form as those of the random phase
approximation (RPA) and agree in detail with the eigenvalue equation that has been derived from the density-matrix version of CKS
theory.  Solutions are to be normalized in the usual way,
according to the conditions (Appendix B),
\begin{equation}
\sum_{ph}(|X_{ph}|^2 -|Y_{ph}|^2) =1.    \label{norm}
\end{equation}
As is well known, two different
non-degenerate solutions of the RPA equations are orthogonal with the same
metric as in (\ref{norm}).

It is important to emphasize what has been accomplished by the calculations
of this section.  With the help of Eq.~(\ref{odden3}), for instance,
we can calculate the off-diagonal matrix elements of the density operator between
the ground state and the first level of excited states.  This can be
applied, for example to the calculation of the corresponding matrix elements
of the electric dipole moment.   We have yet to establish, however, that the eigenvalues $\omega_1$ can be identified
as the excitation energies of the system.  We turn to this task now. 

\section{Excitations as energy differences}

In principle the energy differences can be calculated from the expression
\begin{eqnarray}
\Delta E &\equiv& \sum_{I=0,1}\langle I|\hat{H}|I\rangle -2\langle 0
|\hat{H}|0\rangle  \nonumber \\
&=& E_1 -E_0,    \label{endif2}
\end{eqnarray}
where $E_I$ is the energy of state $I$.
This difference will be evaluated with the aid of Eqs.~(\ref{xc1}), (\ref{xc2}),
and the simplified version of (\ref{fts}). 
The result that we shall establish is 
\begin{eqnarray}
E_1 -E_0 &=& \sum_{p,h}\{(\epsilon_p -\epsilon_h)(|X_{ph}|^2-|Y_{ph}|^2)
+X_{ph}^{\ast}[f_{ph'hp'}X_{p'h'} +f_{pp'hh'}Y_{p'h'}]\nonumber\\
&&+Y_{ph}^{\ast}[f_{hp'ph'}Y_{p'h'} + f_{hh'pp'}X_{p'h'}]\}.  \label{*en}
 \end{eqnarray}
But the right hand side of this equation is easily seen from Eqs.~(\ref{RPA1})
and (\ref{RPA2}) to equal $\omega_1$, provided that we make use of 
Eq.~(\ref{norm}).

It is simplest to evaluate the difference (\ref{endif2}) first for the 
interaction terms. Consider, for instance, the Coulomb difference,
\begin{eqnarray}
\Delta W^c &=& \int d{\bf x}d{\bf x'}\frac{1}{2}\frac{1}{|{\bf x}-{\bf x'}|}
[n^s_{11}({\bf x})n^s_{11}({\bf x'})-n^s_{00}({\bf x})n^s_{00}({\bf x'})
+2n^s_{01}({\bf x})n^s_{10}({\bf x'})]  \nonumber \\
&\approx& \int d{\bf x}d{\bf x'}\frac{1}{|{\bf x}-{\bf x'}|}
\{[n^s_{11}({\bf x})-n^s_{00}({\bf x})]n^s_{00}({\bf x'})
+n^s_{01}({\bf x})n^s_{10}({\bf x'})\},     \nonumber \\
&=& \int d{\bf x}[n^s_{11}({\bf x})-n^s_{00}({\bf x})]v^c({\bf x})+
\int d{\bf x}d{\bf x'}\frac{1}{|{\bf x}-{\bf x'}|}n^s_{01}({\bf x})n^s_{10}({\bf x'})].  \label{cdif}
\end{eqnarray}
To obtain the value exhibited in the first line, we have made the quasi-boson approximation $n_{12}^s=\sqrt{2}n_{01}^s$, which is an expression of the closure approximation referred to in Sec.~II and discussed in more detail below.
The further simplification is made possible by the fact that the difference
$n^s_{11}-n^s_{00}$ (see below) is quadratic in the RPA amplitudes.
The corresponding difference involving the exchange-correlation energy can be written
\begin{eqnarray}
\Delta W^{xc} &=& \int d{\bf x}[n^s_{11}({\bf x})-n^s_{00}({\bf x})]v^{xc}({\bf x})+
\int d{\bf x}d{\bf x'}f(|{\bf x}-{\bf x'}|)n^s_{01}({\bf x})n^s_{10}({\bf x'})],  \label{xcdif}
\end{eqnarray}

Next we see that the second terms of Eqs.~(\ref{cdif})
and (\ref{xcdif}) combine to give 
\begin{eqnarray}
\int d{\bf x}d{\bf x'}f^{eff}(|{\bf x}-{\bf x'}|)n^s_{01}({\bf x})n^s_{10}({\bf x'})
&=& X^{\ast}_{ph}[f_{ph'hp'}X_{p'h'} +f_{pp'hh'}Y_{p'h'}]\nonumber\\ 
&&+Y^{\ast}_{ph}[f_{hp'ph'}Y_{p'h'} + f_{hh'pp'}X_{p'h'}],  \label{xccdif}
\end{eqnarray}
which has been evaluated with the help of Eq.~(\ref{odden3}).  This is already seen to be the interaction terms of Eq.~(\ref{*en}). 

The remaining terms of 
Eqs.~(\ref{cdif}) and (\ref{xcdif}), as well as the contributions arising 
from the kinetic energy and the external potential depend on the value of
\begin{equation}
n^s_{11}({\bf x})-n^s_{00}({\bf x})=\sum_J [\varphi_J^{\ast}({\bf x}1)
\varphi_J({\bf x}1)-\varphi_J^{\ast}({\bf x}0)\varphi_J({\bf x}0)]. \label{endif3}
\end{equation}
To enumerate the states $J$ that contribute to this difference  it is useful to picture
the state $1$ as an elementary {\it boson} excitation, as is done in the 
standard approach to the RPA.  The relations that follow from this assumption
will lead, as we shall see, to a quantitative form of closure 
approximation that is essential to the calculation.  By the notation $1\times 1$, we shall mean a 
double boson excitation with the same boson, whereas by $1\times 1'$ we shall mean
a double excitation with different bosons.  Thus for the amplitudes
$\varphi_J(1)$, we consider the values $J=0h,1h,1\times 1h,1\times 1'h$.  The
contributions from the latter two choices are evaluated in boson 
(closure) approximation as
\begin{eqnarray}
\varphi_{1\times 1h}(1) &=&\sqrt{2}\varphi_1(0),   \label{bos1} \\
\varphi_{1\times 1'h}(1) &=& \varphi_{1'}(0).    \label{bos2}
\end{eqnarray}
For the amplitude $\varphi_J(0)$, the required values are $J=0h,1h,1'h$.  For
the difference (\ref{endif3}), we thus find, suppressing coordinate indices,
\begin{equation}
n^s_{11}-n^s_{00} =\sum_h [\varphi_{0h}^{\ast}(1)\varphi_{0h}(1)+
\varphi_{1h}^{\ast}(0)\varphi_{1h}(0) +\varphi^{\ast}_{1h}(1)\varphi_{1h}(1)
-\varphi_{0h}^{\ast}(0)\varphi_{0h}(0)].     \label{endif4}
\end{equation}

The total contribution of the first two terms of Eq.~(\ref{endif4}) to the 
energy difference under study, obtained by substituting Eqs.~(\ref{rpax}) and
(\ref{rpay}) and applying the result to the sum of single-particle operators
that add up to the KS Hamiltonian $h^s$, is found to be $\sum_{p,h}[\epsilon_p (|X_{ph}|^2
+|Y_{ph}|^2)]$, one of the single-particle terms in Eq.(\ref{*en}).  The evaluation of the remaining terms 
of Eq.~(\ref{endif4}) is carried through by studying the 
normalization conditions, Eq.~(\ref{ks3}). We calculate
\begin{eqnarray}
1&=& \sum_I|\varphi_{0h}(I)|^2   \nonumber \\
&=&  |\varphi_{0h}(0)|^2 +|\varphi_{0h}(1)|^2 +\sum_{I'\neq I}
|\varphi_{0h}(1')|^2,        \label{normzero} \\
1&=& \sum_I|\varphi_{1h}(I)|^2   \nonumber \\
&=& |\varphi_{1h}(1)|^2 +|\varphi_{1h}(0)|^2 +|\varphi_{1h}(1\times 1)|^2
+\sum_{1'\neq 1}|\varphi_{1h}(1\times 1')|^2  \nonumber \\ 
&\approx & |\varphi_{1h}(1)|^2 +|\varphi_{1h}(0)|^2 +2|\varphi_{0h}(1)|^2
+\sum_{1'\neq 1}|\varphi_{0h}(1')|^2,   \label{normone}
\end{eqnarray}
where the last evaluation has made use of the boson approximation expressed by Eqs.~(\ref{bos1}) and (\ref{bos2}).  These equations are satisfied by introducing a renormalization of the KS orbitals
\begin{eqnarray}
\varphi_{0h}({\bf x}0) &=& \varphi_h({\bf x})[1-\frac{1}{2}\sum_p |X_{ph}|^2
-\frac{1}{2}\sum_p\sum_{1'\neq 1}|X_{ph}^{1'}|^2],    \label{renorm0} \\
\varphi_{1h}({\bf x}1) &=&\varphi_h({\bf x})[1-\sum_p |X_{ph}|^2  -\frac{1}{2}
\sum_p |Y_{ph}|^2 -\frac{1}{2}\sum_p\sum_{1'\neq 1}|X_{ph}^{1'}|^2].  \label{renorm1}
\end{eqnarray}
Combining these results and applying them to the last two terms of 
Eq.~(\ref{endif4}), suitably multiplied by the sum of terms that 
comprise $h^s$ leads to 
the final contribution $\sum_{p,h}[-\epsilon_h(|X_{ph}|^2 +|Y_{ph}|^2)]$ to the theorem
stated in Eq.~(\ref{*en}).

\section{Concluding remarks}

In this paper, we have developed an alternative formalism for the study of
excited states within a framework that generalizes the basic ideas of KS theory.
The main novelty in our approach compared to other methods is that the latter work
with a single density, be it the average in the ground state, in an excited state,
an ensemble average, or the average in a suitably chosen time-dependent state.
On the other hand, we arrive at a formalism
involving an entire array of matrix elements of the density operator taken among
a pre-selected set of states.  The application of the variational principle 
for the trace of the Hamiltonian then leads to a generalized KS scheme in terms of
orbitals that depend not only on the coordinate ${\bf x}$, but also on a 
label $I$  of the included states.  We have examined the consequences
of this formalism for the weak-coupling limit.  We did
this by framing a set of assumptions, including 
a closure approximation, in order to identify the most important amplitudes
and their equations
that characterize the ground state and a simple class of excited states that 
are composed of 1p-1h excitations of the ground state.  

In this way, we regained
first the ground-state KS theory and second derived an eigenvalue equation of RPA form. 
By approximating a state-dependent (frequency-dependent) effective interaction
by a state-independent (frequency independent) effective interaction, the
eigenvalue equation became identical to one that was first  derived from the density-matrix 
version of CKS theory \cite{tdf4}.

In our formalism, it is not immediately obvious that the eigenvalues, which originally entered the theory as Lagrange
multipliers in a variational principle, can be identified
with excitation energies of the physical system. We prove that
this is correct identification, in agreement with previous results. The theory also provides the means for the computation
of electromagnetic transition rates from the excited states to the
ground state.  As formulated, the theory described in this paper
can be extended to improve the approximations made for $1qp-1qh$
states, as well as to study more complicated states, for example,
of $2qp-2qh$ character.

\section*{Acknowledgment}

One of the authors (AK) is grateful to the Humboldt Foundation for support of this work and to his co-author for his hospitality.

\appendix

\section{Relation of weak-coupling limit to time-dependent
density functional theory}

In this section, we shall connect the linearized RPA equations (\ref{RPA1})
and (\ref{RPA2}) with a corresponding linearized approximation to
TDDFT.  We start with TDDFT in density-matrix form
\begin{eqnarray}
i\frac{d\rho^s}{dt} &=& [(\tau + v^s(t)), \rho^s], \label{a1} \\
\rho^s({\bf x}t,{\bf x'}t) &=& \sum_h \varphi_h({\bf x}t)
\varphi^{\ast}_h({\bf x'}t),    \label{a2}  \\
v^s({\bf x}t) &=& \frac{\delta}{\delta n({\bf x}t)}(V(t) +W(t) +T(t)
-T^s(t)). \label{a3}
\end{eqnarray}
Here $\varphi({\bf x}t)$ are the $N$ instantaneous eigenfunctions of
$\tau + v^s(t)$ of lowest energy, defining a time-dependent Slater
determinant whose kinetic energy is $T^s(t)$, and $V(t)$, for example,
is the expectation value of $\hat{V}$ in the time-dependent wave-function
$|\Psi(t)\rangle$.

We are interested in the physical situation where the time-dependence
of the state vector arises
not from an explicitly time-dependent external field but from the fact that
initially the state vector is a superposition of the ground state
(predominately) and a small amplitude for one of the excited states.
We thus assume that
\begin{eqnarray}
\rho^s({\bf x}t,{\bf x'}t) &=& \rho^0({\bf x,x'})
+[\rho^1({\bf x,x'})\exp(-i\omega t)  +\;\;{\rm c.c.}], \label{a4} \\
\rho^1({\bf x,x'}) &=& \sum_{ph}[X_{ph}\varphi_p({\bf x})
\varphi^{\ast}_h({\bf x'})  +Y_{ph}\varphi_h({\bf x})
\varphi^{\ast}_p({\bf x'})].    \label{a5}
\end{eqnarray}
In (\ref{a4}) and below the superscript $0$ identifies quantities associated
with the KS ground-state theory. 

What follows now is close to a standard derivation of the RPA.  We insert
(\ref{a4}) and (\ref{a5}) into (\ref{a1}) and, considering the amplitudes
$X$ and $Y$ as first order quantities, we expand to first order.
For this purpose, we need the expansion,
\begin{eqnarray}
v^s({\bf x}t) &=& v^0({\bf x}) +\int d{\bf x'}f({\bf x,x'})n^1({\bf x'}),
\label{a6} \\
f({\bf x,x'}) &=& \frac{\delta v^0({\bf x})}{\delta n^0({\bf x'})},
\label{a7} \\
n^1({\bf x}) &=& \rho^1({\bf x,x}).    \label{a8}
\end{eqnarray}
In Eqs.~(\ref{a6}) and (\ref{a7}), we have already made the adiabatic approximation by ignoring the time dependence of $f$.
As a consequence, the quantity called $f$ in this appendix can be identified with the quantity $f^{eff}$ of the text.
From the zero order term, we regain the KS theory for the ground state.
From the first order terms proportional to $\exp(-i\omega t)$, for example,
we find
\begin{eqnarray}
\omega\rho^1({\bf x,x'}) &=& [(\tau + v^0),\rho^1]({\bf x,x'}) 
+\int d{\bf x''} [\frac{\delta v^0}{\delta n({\bf x''})},\rho^0]({\bf x,x'})
n^1({x''}).  \label{a9}
\end{eqnarray}

Taking, in turn, the $qph$ and $qhp$ matrix elements of (\ref{a9}), 
we find the familiar equations
\begin{eqnarray}
{}[\epsilon_{h} -\epsilon_{p} +\omega]X_{ph}&=&f_{ph'hp'}X_{p'h'}
 +f_{pp'hh'}Y_{p'h'},   \label{a11} \\
{}[\epsilon_{h} -\epsilon_{p} -\omega]Y_{ph}&=&f_{hp'ph'}Y_{p'h'}
 +f_{hh'pp'}X_{p'h'}.   \label{a12}
\end{eqnarray}

\section{RPA normalization condition}

We define mode operators, $a_a$, for the field $\hat{\psi}({\bf x})$ by expanding in terms of the KS modes,
\begin{equation}
\hat{\psi}({\bf x}) = \sum_a a_a\varphi_a({\bf x}), \label{b1}
\end{equation}
$a=\{h,p\}$.  From the commutation relations for quasiparticle-quasihole pairs,
\begin{eqnarray}
[a_{h}^{\dag}a_{p},a_{p'}^{\dag}a_{h'}] &=&\delta_{hh'}\delta_{pp'} 
 -\delta_{hh'}a_{p'}^{\dag}a_p -\delta_{pp'}a_{h'}a_h^{\dag}, \label{b2}
\end{eqnarray}
we obtain an approximate sum rule by taking the expectation value in the
state $|0\rangle$, introducing a complete set of intermediate states
$|i\rangle$, and retaining only the first term on the right hand side
(on the justified assumption that, for instance, $\langle 0|a_p^{\dag}
a_p'|0\rangle$ is, on the average small compared to unity).  With the
definitions
\begin{eqnarray}
\xi_{ph}^i &=& \langle 0|a_h^{\dag}a_p|i\rangle,   \label{b3} \\
\eta_{ph}^i &=& \langle 0|a_p^{\dag}a_h|i\rangle,   \label{b4}
\end{eqnarray}
we have
\begin{equation}
\sum_i [\xi_{ph}^i \xi_{p'h'}^{i\ast} - \eta_{p'h'}^i \eta_{ph}^{i\ast}]
 =\delta_{pp'}\delta_{hh'}.  \label{b5}
\end{equation}

We would like to identify the quantities $\xi$ and $\eta$ with the quantities
$X$ and $Y$, where the latter satisfy Eqs. (\ref{RPA1}) and (\ref{RPA2}).
Equation (\ref{b5}) would then constitute the completeness relation for the
solutions of these equations, and as is well-known, a completeness relation
and orthogonality of solutions with the corresponding metric implies the
normalization condition Eq.~(\ref{norm}).  Toward this end,
we consider two different evaluations of $\langle 0|\hat{\psi}^{\dag}
({\bf x})\hat{\psi}({\bf x})|i\rangle =n_{i0}({\bf x})$. On the one hand
we have in an approximate evaluation based on the physical picture,
\begin{eqnarray}
n_{i0}({\bf x}) &=& \sum_{ab} \varphi_a^{\ast}({\bf x})\varphi_b({\bf x})
 \langle 0|a_a^{\dag}a_b|i\rangle    \nonumber \\
& \cong & \sum_{ph}[\varphi_p^{\ast}({\bf x})\varphi_h({\bf x})
\langle 0|a_p^{\dag}a_h|i\rangle]  
 +\varphi_h^{\ast}({\bf x})\varphi_p({\bf x})
\langle 0|a_h^{\dag}a_p|i\rangle.  \label{b6}
\end{eqnarray}
On the other hand, from the generalized KS mapping $n_{i0}\rightarrow
n_{i0}^s$ and Eq.~(\ref{odden3}, we have 
\begin{equation}
n_{i0}({\bf x}) = \sum_{ph}[\varphi_p^{\ast}({\bf x})\varphi_h({\bf x})
Y_{ph}^i + \varphi_h^{\ast}({\bf x})\varphi_p({\bf x})X_{ph}^i] .
\label{b7}
\end{equation}
The identifications $\xi=X$ and $\eta=Y$ are
consistent with these equations.


\begin{thebibliography} {99}

\bibitem{KD}
P.\ Hohenberg and W.\ Kohn, Phys.\ Rev.\ {\bf 136B}, 864 (1964).
\bibitem {KS}
W.\ Kohn and L.\ J.\ Sham, Phys.\ Rev.\ {\bf 140A}, 1133 (1965).
\bibitem {PY}
R.\ G.\ Parr and W.\ Yang, {\it Density-Functional Theory of Atoms and
Molecules}, (Oxford U.\ Press, New York, 1989).
\bibitem {DG} 
R.\ M.\ Dreizler and E.\ K.\ U.\ Gross, {\it Density Functional
Theory, An Approach to the Quantum Many-Body Problem}, 
(Springer-Verlag, Berlin, 1990).
\bibitem{HM}
A.\ Holas and M.\ H.\ March, in {\em Topics in Current Chemistry}, Vol.\ 180,
ed.\ R.\ F.\ Nalewajski (Springer, Berlin, 1996), p.\ 57.
\bibitem{tr1}
R.\ Courant and D.\ Hilbert, {\em Methods of Mathematical Physics}
(Interscience, New York, 1965), Vol.\ 1, p.\ 459.
\bibitem{tr2}
A.\ K.\ Theophilou, J.\ Phys.\ C {\bf 12}, 5419 (1979).
\bibitem{tr3}
E.\ K.\ U.\ Gross, L.\ N.\ Oliveira, and W.\ Kohn, Phys.\ Rev.\ A {\bf 37},
2805 (1988); {\bf 37}, 2809 (1988); {\bf 37}, 2821 1988).
\bibitem{tdf1}
E.\ K.\ U.\ Gross, J.\ F.\ Dobson, and M.\ Petersilka, in
{\em Topics in Current Chemistry}, Vol.\ 181, ed.\ R.\ F.\ Nalewajski
(Springer, Berlin, 1996), p.\ 81.
\bibitem{tdf2}
M.\ Petersilka, U.\ J.\ Gossmann, and E.\ K.\ U.\ Gross, Phys.\ Rev.\ Lett.\
{\bf 76}, 1212 (1996).
\bibitem{tdf3}
J.\ F.\ Dobson, M.\ J.\ Buner, and E.\ K.\ U.\ Gross, Phys.\ Rev.\ Lett.\
{\bf 79}, 1905 (1997).
\bibitem{tdf4}
M.\ E.\ Casida, in {\em Recent Advances in Density Functional Methods,
Part I}, ed.\ D.\ P.\ Chong (World Scientific, Singapore, 1995), p.\ 155.
\bibitem{tdf5}
M.\ e.\ Casida {\em et al}, in {\em Nonlinear Optical Materials}, ed.\
S.\ P.\ Karna and A.\ T.\ Yeates (American Chemical Society, Washington),
p.\ 145.
\bibitem{tdf6}
C.\ Jamorski, M.\ E.\ Casida, and D.\ R.\ Salahub, J.\ Chem.\ Phys.\
{\bf 104}, 5134 (1996).
\bibitem{GHR}
A.\ G{\"o}rling, H.\ Heinze, S.\ Ph.\ Ruzamkin, M.\ Stauffer,
and N. R{\"o}sch, J.\ Chem.\ Phys.\ {\bf 110}, 2785 (1999).
\bibitem{BA1}
R.\ Bauernschmitt and R.\ Ahlrichs, Chem.\ Phys.\ Lett.\ {\bf 256}, 454 (1996).
\bibitem{BA2}
R.\ Bauernschmitt, M.\ H\"aser, O.\ Treutler, and R.\ Ahlrichs, Chem.\ Phys.\
Lett.\ {\bf 264}, 573 (1997).
\bibitem{HG1}
S.\ Hirata and M.\ Head-Gordon, Chem.\ Phys.\ Lett.\ {\bf 302}, 375 (1999).
\bibitem{HG2}
S.\ Hirata and M.\ Head-Gordon, Chem.\ Phys.\ Lett.\ {\bf 314}, 291 (1999).
\bibitem{WSF1}
K.\ B.\ Wiberg, R.\ E.\ Stratmann, and M.\ J.\ Frisch, Chem.\ Phys.\ Lett.\
{\bf 297}, 60 (1998).
\bibitem{WSF2}
R.\ E.\ Stratmann, G.\ E.\ Scuseria, and M.\ J.\ Frisch, J.\ Chem.\ Phys.\
{\bf 109}, 8218 (1998).
\bibitem{AG}
A.\ G{\"o}rling, Phys.\ Rev.\ A {\bf 59}, 3359 (1999).
\bibitem{LG}
M.\ Levy and A.\ Nagy, Phys.\ Rev.\ Lett.\ {\bf 83}, 4361 (1999)
\bibitem{GS}
X.\ Gonze and M.\ Scheffler, Phys.\ Rev.\ Lett.\ {\bf 82}, 4416 (1999).
\bibitem{TH}
D.\ J.\ Tozer and N.\ C.\ Handy, J.\ Chem.\ Phys.\ {\bf 109}, 10180 (1998).
\bibitem{kk1}
A.\ K.\ Kerman and A.\ Klein, Phys.\ Rev.\ {\bf 132}, 1326 (1963).
\bibitem{kk2}
G.\ Do Dang {\em et al}, Nucl.\ Phys.\ {\bf A114}, 501 (1968).
\bibitem{kk3}
A.\ Klein, Phys.\ Rev.\ {\bf C30}, 1680 (1984).
\bibitem{no}
A.\ Klein and R.\ M.\ Dreizler, Phys.\ Rev.\ {\bf A58}, 1581 (1998).



\end{thebibliography}
\end{document}